# On the pricing of currency options under variance gamma process

Azwar Abdulsalam[*]    Gowri Jayprakash[†]    Abhijeet Chandra[‡§]

## Abstract

The pricing of currency options is largely dependent on the dynamic relationship between a pair of currencies. Typically, the pricing of options with payoffs dependent on multi-assets becomes tricky for reasons such as the non-Gaussian distribution of financial variable and non-linear macroeconomic relations between these markets. We study the options based on the currency pair US dollar and Indian rupee (USD-INR) and test several pricing formulas to evaluate the performance under different volatility regimes. We show the performance of the variance gamma and the symmetric variance gamma models during different volatility periods as well as for different moneyness, in comparison to the modified Black-Scholes model. In all cases, variance gamma model outperforms Black-Scholes. This can be attributed to the control of kurtosis and skewness of the distribution that is possible using the variance gamma model. Our findings support the superiority of variance gamma process of currency option pricing in better risk management strategies.

*Keywords*: Currency options, Variance gamma process, Risk management.

*JEL Codes*: F31, G15, G32

---

[*] Department of Electronics and Electrical Communication, IIT Kharagpur, India
[†] Department of Ocean Engineering and Naval Architecture, IIT Kharagpur, India
[‡] Vinod Gupta School of Management, Indian Institute of Technology Kharagpur, India
[§] Corresponding author. Email: abhijeet@vgsom.iitkgp.ac.in. Tel: +91-3222-281780.

# On the pricing of currency options under variance gamma process

**Introduction**

Currency options are an important instrument for traders to hedge against possible fluctuations in exchange rates and at the same time have high gain potentials while taking on limited risks. Since the introduction of option trading on USD-INR in NSE there has been an exponential growth in the volume of contracts traded from a monthly average of 139,296 contracts in 2010 to 54,078,805 in 2019. Thus it is imperative we have a robust model for pricing these contracts. In this paper we have carried out an empirical study to see the performance of variance gamma and the symmetric variance gamma model in pricing of USD-INR during different volatility periods as well as for different moneyness. The results were then compared with that obtained from the modified Black-Scholes. In all cases Variance gamma model was found to out-perform Black-Scholes. This can be attributed to the control of kurtosis and skewness of the distribution that is possible using the variance gamma model. Section II provide gives a brief overview of the different models we have considered and section III explores the empirical and statistical results from these models.

**Methodology**

1. *Black-Scholes Garman–Kohlhagen Model:*

The modified Black-Scholes Model was proposed by Garman and Kohlhagen in 1983 to take into account the difference in interest rates between two countries while valuing currency options. It is a pure diffusion process and does not involve any jump component. The modified Black-Scholes formula is:

$$C_{BS} = S_o N(d_1) e^{-r_f T} - K N(d_2) e^{-r_d T} \qquad (1)$$

Where $d_1 = \dfrac{\ln\left(\frac{S}{k}\right) + (r_d - r_f + \sigma^2/2)T}{\sigma\sqrt{T}}$

$d_2 = d_1 - \sigma\sqrt{T}$

$N(.)$ represents the cumulative normal distribution $S_o$ is the current exchange rate, $r_f$ is the foreign interest rate $r_d$ is the domestic risk free interest rate and T is the time to expiration of the contract.

2. *Variance Gamma Model:*

The variance gamma model was proposed by Carr and Madan in 1992. Unlike most models in the literature Variance is a pure jump process with no continuous Martingale component [1]. Variance gamma process $(X(t; \sigma, \nu, \theta))$ is basically geometric Brownian motion evaluated at random time having a gamma distribution i.e.

$$X(t;\sigma,\nu,\theta) = b(T_t;\theta,\sigma) \qquad (2)$$

Where $b(T_t;\theta,\sigma)$ represents a Brownian motion having drift $\theta$ and variance $\sigma$ and $T_t$ is the value of a Gamma process $\gamma(t;\mu,\nu)$ at time t with mean $\mu = 1$ and variance $\nu$. Thus the variance Gamma process involves two steps 1) picking a value of $T_t$ 2) evaluating the Brownian motion $b(T_t;\theta,\sigma)$. The conditional probability distribution of $X$ is given by:

$$f(X|T_t = g) = \frac{1}{\sigma\sqrt{2\pi g}} \exp\left(\frac{x-\theta g}{2\sigma^2 g}\right) \qquad (3)$$

Thus the unconditional probability distribution is given by integrating over all possible values of $T_t$ i.e.

$$f_{X(t)}(X) = \int_0^\infty f(X|T_t) * \gamma(t;\mu,\nu) \qquad (4.1)$$

$$= \int_0^\infty \frac{1}{\sigma\sqrt{2\pi g}} \exp\left(\frac{x-\theta g}{2\sigma^2 g}\right) \frac{g^{\frac{t}{\nu}-1} \exp(-\frac{g}{\nu})}{\nu^{\frac{t}{\nu}} * \Gamma(\frac{t}{\nu})} dg \qquad (4.2)$$

The kurtosis of the distribution is controlled by the factor $\nu$ and the skewness is captured by the factor $\theta$. If we assume the stock prices to follow a Variance Gamma process we have

$$S(t) = S(0)\exp(X(t;\sigma,\nu,\theta) + \omega t + rt) \qquad (5)$$

Where r is the risk free interest rate and $\omega$ is correction term required to make the overall measure martingale and is given by

$$\omega = \frac{t}{v}\log\left(1 - \theta v - \frac{\sigma^2 v}{2}\right) \tag{6}$$

The distribution of stock price is thus dependent on the realization of the random variable $T_t$ which then is given by a log-normal distribution. For the unconditional probability distribution we have to integrate as done earlier in (4) which then results in the following expression for the log returns $\left(z = \ln\left(\frac{S(t)}{S(0)}\right)\right)$:

$$\frac{2\exp(\frac{\theta x}{\sigma^2})}{v^{\frac{t}{v}}\sqrt{2\pi}\sigma\Gamma(\frac{t}{v})} * \left(\frac{x^2}{\frac{2\sigma^2}{v}+\theta^2}\right)^{\frac{t}{2v}-\frac{1}{4}} * K_{\frac{t}{v}-\frac{1}{2}}\left(\frac{1}{\sigma^2}\right)\sqrt{x^2(\frac{2\sigma^2}{v}+\theta^2)} \tag{7}$$

Where K is the modified Bessel function of the second order and $x$ is given by

$$x = z - mt - \frac{t}{v}\log\left(1 - \theta v - \frac{\sigma^2 v}{2}\right) \tag{8}$$

Once the paramaters of the Variance gamma are selected i.e. $v, \theta$ and $\sigma$ we can price an european option as follows:

$$C_{VG} = e^{-rT}\mathbb{E}(S(0)\exp(X(t;\sigma,v,\theta) + \omega t + rt)) \tag{9}$$

Upon evaluating (9.1) on obtains the following closed form solution for the pricing of european currency option[2]:

$$C_{VG}(t) = S(t)e^{-r_f T}\psi\left(d\sqrt{\frac{1-c_1}{v}}, (\alpha+s)\sqrt{\frac{v}{1-c_1}}, \frac{T}{v}\right) -$$

$$Ke^{-r_f T}\psi\left(d\sqrt{\frac{1-c_2}{v}}, \alpha s\sqrt{\frac{v}{1-c_1}}, \frac{T}{v}\right) \quad (10)$$

$$\text{Where } d = \frac{1}{s}\left[\ln\left(\frac{S(t)}{K}\right) + (r_d - r_f)T + \frac{T}{v}\ln\left(\frac{1-c_1}{1-c_2}\right)\right]$$

$$c_1 = v(\alpha+s)^2/2$$

$$c_2 = v\alpha^2/2$$

$$\alpha = -\theta s/\sigma^2$$

$$s = \frac{\sigma}{\sqrt{1 + \left(\frac{\theta}{\sigma}\right)^2 \frac{v}{2}}}$$

The function $\psi$ is given in the appendix and is expressed in terms of modifed bessel function of second order and hypergeomteric function.

**Empirical Evidence**

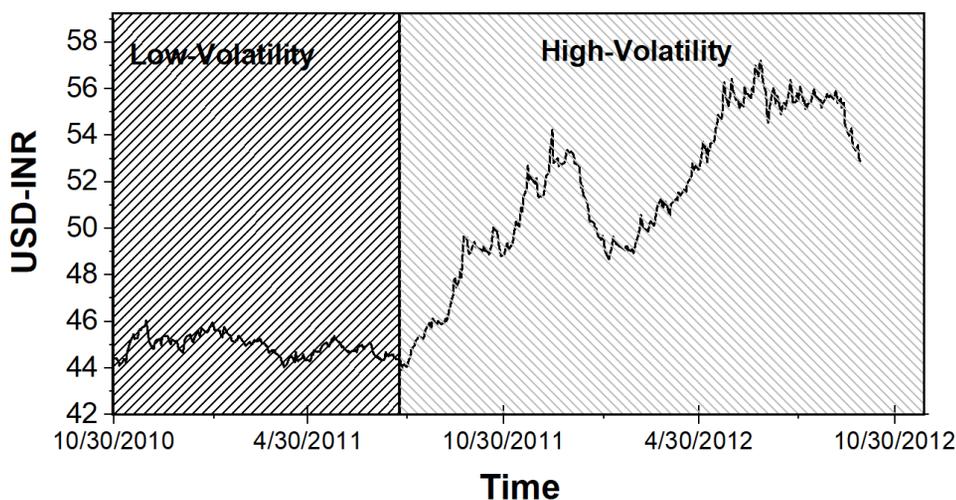

Figure1: USD-INR exchange rate for the time period Nov 1st 2010 to Sep 28th 2012 divided into two periods based on the volatility. Low volatility period is from Nov 1st 2010 to July 28th 2011 and high volatility is from July 28th 2011 to Sep 28th 2012.

The data we have considered is the USD-INR option price for call options obtained from the National Stock Exchange-NSE for the duration of Nov 1st 2010 to Sep 28th 2012. We have considered contracts having volume greater than 100 in order to avoid taking into consideration illiquid contracts which may not be priced appropriately. We have a total of 7312 contracts which averages to approximately 80 contracts a week. To further understand the performance of the models in periods of high volatility and low volatility we have divided our timeline into two periods Nov 1st 2010 to July 28th 2011 as low volatility period and from July 28th 2011 to Sep 28th 2012 as high volatility period as shown in Fig.1. The option data is then further divided based on the moneyness as at the money (ATM; S/K>0.95 and S/K<1.05) in the money (ITM; S/K>1.05) and out of the money (S/K<0.95) for each period as shown in Table 1.

**TABLE 1**

| Moneyness | Low Volatility Period | High Volatility Period |
|:---:|:---:|:---:|
| ITM | 18 | 675 |
| ATM | 2565 | 5909 |
| OTM | 48 | 648 |
| Total | 2631 | 7232 |

1. *Historical data parameters:*

In table II we have shown the parameters for each distribution which fits with the historical data for the two time periods we have considered. For the case of Black-Scholes volatility ($\sigma$) was calculated by taking the variance of the historical data. For the case of Variance Gamma and symmetric Variance Gamma was calculated by making an initial approximation using the moment fucntions (11.1)-(11.4). These values were then used as initial guesses to fit the historical data with (7) using Nelder-Mead optimization algorithm.

$$\mathbb{E}(X_t) = c + \theta \tag{11.1}$$

$$Var(X_t) = \sigma^2 \tag{11.2}$$

$$\mathbb{E}(X_t - \mathbb{E}(X_t))^3 = 3\sigma^2 \theta v \tag{11.3}$$

$$\mathbb{E}(X_t - \mathbb{E}(X_t))^4 = 3\sigma^4(1+v) \tag{11.4}$$

## TABLE 2

| Parameters | Black-Scholes | Variance Gamma |
|---|---|---|
| $\sigma$ (HV) | 0.1039 | 0.1044 |
| (LV) | (0.054) | (0.0545) |
| $\theta$ (HV) | - | -0.00118 |
| (LV) |  | (0.00682) |
| $\nu$ (HV) | - | 0.211 |
| (LV) |  | (0.083) |

Above Table shows the distribution parameters for the three models i.e. Black Scholes, Symmetric Variance Gamma and Variance Gamma. These values have been obtained by calibrating the model distribution with the daily log returns of the USD-INR exchange rate in the interval Dec 30th 2010 to April 30th 2012.

The volatility($\sigma$) for the low volatility time period was found to be 0.054 for Black-Scholes and 0.0544 and 0.0545 for the case of symmetric Variance Gamma and Variance Gamma respectively. For high volatility period it was found to be 0.1039 for the case of Black-Scholes and 0.1044 for both the symmetric Variance Gamma and Variance Gamma models. The much higher value of $\sigma$ obtained for all models in the high volatility period compared to the low volatility period is consistent with the time period division. The kurtosis of Variance Gamma and symmetric Variance Gamma given by $\kappa = 3(1 + \nu)$ gives a value of 3.606 and 3.379 in the case of symmetric VG for low and high volatile periods respectively whereas for VG it was observed to be 3.633 and 3.249 for low and high volatile periods respectively. Skewness of the distribution specified by $\theta$ indicates a positive skewness for the low volatile period however for the high volatile period it is found to have a negative skewness.

2. *Risk Neutral parameters:*

In Table III is shown the mean weekly risk neutral parameters of the different models and their standard deviation. The weekly parameters have been calculated by minimizing the log-likelihood function $\sum_{i=0}^{M} |\log(\mathbb{C}_{Model_i}(\theta_1, \theta_2, \theta_3, \dots)) - \log(\mathbb{C}_{Market_i})|$, where $\mathbb{C}_{Model}(\theta_1, \theta_2, \theta_3, \dots)$ is the option price calculated from the respective model and $\theta_1, \theta_2, \theta_3, \dots$ are the various model parameters for which $f$ is minimum. The optimization algorithm used is Nelder-Mead and for the initial guess we used the parameters obtained from the historical values shown in Table II.

TABLE 3

| Model | | Mean | Standard Deviation | Maximum | Minimum |
|---|---|---|---|---|---|
| Black-Scholes: $\sigma$ | | 0.08275 | 0.0216 | 0.1233 | 0.0455 |
| Variance Gamma: | $\nu$ | 0.099 | 0.06 | 0.3471 | 0.0091 |
| | $\theta$ | 0.0026 | 0.0048 | 0.0188 | -0.0028 |
| | $\sigma$ | 0.116 | 0.00517 | 0.8651 | 0.002 |

The weekly risk neutral parameters obtained are tabulated and shown in table 3 along with their statistical distribution.

**Out of Sample Performance:**

In order to the test the out of sample performance of the different models we have used one weeks on implied model parameters to predict the option price for the next week. This is in line with the previous works of Madan (1998). Below shown is the statistics of the percentage error of the two models.

**TABLE 4**

| Model | Mean | Standard Deviation | Maximum | Minimum |
|---|---|---|---|---|
| Black-Scholes | 0.244 | 0.0786 | 0.511 | 0.11 |
| LV period | 0.191 | 0.052 | 0.301 | 0.11 |
| HV period | 0.272 | 0.0707 | 0.511 | 0.158 |
| Variance Gamma | 0.05848 | 0.03041 | 0.156 | 0.0012 |
| LV period | 0.0446 | 0.0334 | 0.149 | 0.016 |
| HV period | 0.068 | 0.024 | 0.156 | 0.034 |

In table 4 above is shown the mean absolute relative error i.e. (MAPE) = $\sum_{i=1}^{n}(\mathbb{C}_{Market_i} - \mathbb{C}_{Model_i})/n$. It is seen that overall mean of MAPE of Variance Gamma model with 0.0584 is much lower than that of Black Scholes's 0.244. This is evidence to the robustness of the VG model compared to Black-Scholes. In order to understand the performance of the two models in different cases such as for based on the maturity of the contract as well as based on moneyness we have further compiled its performance in table 5 and 6. In table 5 is shown performance of the two models based on the maturity period.

**TABLE 5**

| Parameter | Mean | Standard Deviation | Maximum | Minimum |
|---|---|---|---|---|
| Black-Scholes <30days | 0.123 | 0.076 | 0.307 | 0.0012 |
| Black-Scholes >30days and <60 days | 0.081 | 0.053 | 0.2876 | 0.023 |
| Black-Scholes >60days | 0.038 | 0.041 | 0.238 | 0.015 |
| VG Short Term <30 days | 0.035 | 0.023 | 0.125 | 0.0015 |
| VG Med Term >30 and <60 days | 0.015 | 0.031 | 0.301 | 0.0012 |
| VG Long Term >60 | 0.06 | 0.063 | 0.623 | 0.0018 |

**TABLE 6**

| Parameter | Mean | Standard Deviation | Maximum | Minimum |
|---|---|---|---|---|
| Black-Scholes ITM | 0.0034 | 0.0039 | 0.017 | 0.0001 |
| Black-Scholes ATM | 0.223 | 0.066 | 0.45 | 0.1014 |
| Black-Scholes OTM | 0.014 | 0.020 | 0.097 | 0.015 |
| VG Short Term ITM | 0.0118 | 0.021 | 0.121 | 0.003 |
| VG Med Term ATM | 0.0458 | 0.0271 | 0.116 | 0.012 |
| VG Long Term OTM | 0.011 | 0.021 | 0.1215 | 0.0013 |

**Conclusions**

In this paper we have carried out an empirical study to see the performance of variance gamma and the symmetric variance gamma model in pricing of USD-INR during different volatility periods as well as for different moneyness. The results were then compared with that obtained from the modified Black-Scholes. In all cases Variance gamma model was found to out-perform Black-Scholes. This can be attributed to the control of kurtosis and skewness of the distribution that is possible using the variance gamma model.